\documentclass[runningheads]{llncs}
\usepackage{amsmath,amsfonts}
\usepackage{graphicx}
\usepackage{booktabs}
\usepackage{adjustbox}
\usepackage{bbding}
\usepackage[marginal]{footmisc}

\begin{document}

\title{Three-Dimensional Medical Image Fusion with Deformable Cross-Attention}
%
%
\author{Lin Liu\inst{1,2} \and
Xinxin Fan\inst{1,2}\and
Chulong Zhang\inst{1}\and
Jingjing Dai\inst{1}\and
Yaoqin Xie\inst{1}\and
Xiaokun Liang\inst{1(}\Envelope\inst{)}}
\authorrunning{F. Author et al.}
%
\institute{Shenzhen Institute of Advanced Technology, Chinese Academy of Sciences, Shenzhen 518055, China\\
\email{xk.liang@siat.ac.cn}\\
\and
University of Chinese Academy of Sciences, Beijing 100049, China\\}

\maketitle              
\footnote{First Author and Second Author contribute equally to this work.\\}
\begin{abstract}
Multimodal medical image fusion plays an instrumental role in several areas of medical image processing, particularly in disease recognition and tumor detection. Traditional fusion methods tend to process each modality independently before combining the features and reconstructing the fusion image. However, this approach often neglects the fundamental commonalities and disparities between multimodal information. Furthermore, the prevailing methodologies are largely confined to fusing two-dimensional (2D) medical image slices, leading to a lack of contextual supervision in the fusion images and subsequently, a decreased information yield for physicians relative to three-dimensional (3D) images. In this study, we introduce an innovative unsupervised feature mutual learning fusion network designed to rectify these limitations. Our approach incorporates a Deformable Cross Feature Blend (DCFB) module that facilitates the dual modalities in discerning their respective similarities and differences. We have applied our model to the fusion of 3D MRI and PET images obtained from 660 patients in the Alzheimer's Disease Neuroimaging Initiative (ADNI) dataset. Through the application of the DCFB module, our network generates high-quality MRI-PET fusion images. Experimental results demonstrate that our method surpasses traditional 2D image fusion methods in performance metrics such as Peak Signal to Noise Ratio (PSNR) and Structural Similarity Index Measure (SSIM). Importantly, the capacity of our method to fuse 3D images enhances the information available to physicians and researchers, thus marking a significant step forward in the field. The code will soon be available online.

\end{abstract}
\section{Introduction}

Multimodal image fusion represents a crucial task in the field of medical image analysis. By integrating information from diverse imaging modalities, image fusion leverages the complementary characteristics of each technique. Each modality inherently focuses on distinct physiological or pathological traits. Hence, an effective fusion of these attributes can yield a more holistic and intricate image, thereby easing and enhancing physicians' decision-making processes during diagnosis and treatment \cite{james2014medical}.

For instance, the significance of Magnetic Resonance Imaging (MRI) and Positron Emission Tomography (PET) in image fusion is particularly noteworthy. MRI provides excellent soft tissue contrast and high-resolution anatomical structure information, while PET can present images of metabolic activity and biological processes \cite{huang2020review}. The resultant image, derived from the fusion of MRI and PET, provides a comprehensive perspective, encompassing detailed anatomical structures alongside metabolic function data. This amalgamation plays a vital role in early disease or tumor detection and localization and significantly influences subsequent treatment strategies \cite{yin2018medical}.

Traditionally, medical image fusion primarily utilizes multi-scale transformations in the transform domain, which typically comprises three steps. First, the images from each modality are subjected to specific transformations, such as wavelet \cite{singh2013multimodal,chabi2013efficient,talbi2018predator} or pyramid transformations \cite{du2016union,burt1993enhanced}, yielding a series of multi-scale images. Then, these multi-scale images at the same scale level are analyzed and selected to retain the most representative information. Finally, through an inverse transformation, these multi-scale images are amalgamated into a novel image. Beyond these methods, sparse representation has also been applied in image fusion \cite{yang2012pixel,liu2015simultaneous,liu2016image}.

Nevertheless, image fusion tasks face significant hurdles, primarily due to the absence of a "gold standard" or "ground truth" that could encapsulate all modality information—ideally, a comprehensive reference image. Traditional fusion techniques often falter when dealing with high-dimensional data, particularly when encountering noise and complex data distributions across modalities. Moreover, the design of fusion rules remains manual, resulting in suboptimal generalization and unresolved semantic conflicts between different modality images \cite{hill2016perceptual}. Deep learning, with its inherent capacity for automatic feature learning and multi-layer abstraction, is poised to mitigate these challenges, enabling more accurate and interpretable image fusion.

Despite the progress, existing medical image fusion techniques primarily focus on two-dimensional (2D) slice fusion, which presents clear limitations. Medical images are predominantly three-dimensional (3D) signals, and 2D fusion approaches often neglect inter-slice context information. This oversight leads to a degree of misinterpretation of spatial relationships crucial for decoding complex anatomical structures. It is in this context that the potential benefits of 3D fusion become salient. By incorporating all 3D of space, 3D fusion can deliver more accurate localization information—a critical advantage in applications requiring precision, such as surgical planning and radiation therapy. Equally important, 3D fusion affords a panoramic view, enabling physicians to inspect and analyze anatomical structures and physiological functions from any perspective, thereby acquiring more comprehensive and nuanced information.

In this paper, we focus on the fusion of PET and MRI medical images, although the proposed methodology is generalizable to other imaging modalities. Our contributions can be delineated as follows: 1) We break new ground by applying a deep learning-based framework for 3D medical image fusion. 2) We introduce a Deformable Cross-Feature Fusion module that adjusts the correspondence information between the two modalities via Positional Relationship Estimation (PRE) and cross-fuses the features of the two modalities, thus facilitating image feature fusion. 3) We evaluate our approach on publicly available datasets, and our method yields state-of-the-art results quantitatively, based on Peak Signal-to-Noise Ratio (PSNR) and Structural Similarity Index Measure (SSIM). Qualitatively, our technique, even within the same 2D slice, surpasses baseline methods focusing on 2D fusion by not only retaining ample PET information but also integrating MRI structural data. The structural information discerned by our approach aligns more closely with the original MRI image.

\section{Methods}

\subsection{Overview of Proposed Method} 
\begin{figure}
	\centering
	\begin{adjustbox}{width= 1.1\textwidth,center}
		\includegraphics{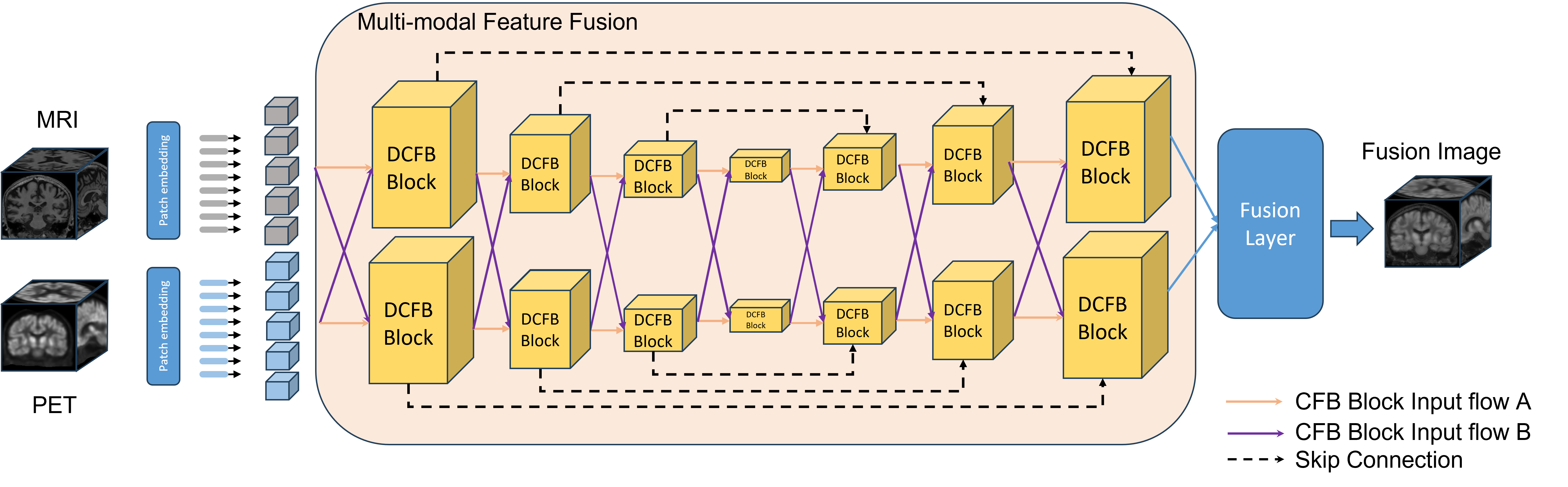} 
	\end{adjustbox}
	\caption{Overall architecture of the proposed model comprising Patch Embedding and a fully mirrored symmetric Multi-modal Feature Fusion module, followed by a Fusion Layer consisting of fully convolutional operations.}
	\label{fig:overview}
\end{figure}

In this study, we utilized a U-shaped architecture for MRI-PET image fusion, shown in Fig.~\ref{fig:overview}. The network employed a dual-channel input, with MRI and PET 3D image information fed separately. To reduce the parameter count of the 3D network, we applied Patch Embedding using DC2Fusion to both inputs. The resulting patch images were then passed through the Multi-modal Feature Fusion (MMFF) module. MMFF consisted entirely of Cross Fusion Blend (CFB) blocks at different scale levels. Notably, MMFF exhibited a fully symmetrical network structure, allowing the MRI branch to learn PET image information and the PET branch to acquire MRI image information. This symmetrical information interaction was facilitated by the CFB block, which had two input streams: input flow A and input flow B. In the MRI branch, MRI image information served as input flow A, while PET image information served as input flow B. The CFB block enabled the interaction and fusion of features between the two inputs. Consequently, the MRI branch adjusted its own features after perceiving certain characteristics in the PET image and outputted MRI image feature information. When the CFB block operated in the PET branch, input flow A and input flow B reversed their order compared to the MRI branch, enabling the adjustment of PET image features. After passing through the MMFF module, all branches entered a Fusion Layer composed of convolutions to merge the features and reconstruct the Fusion image.

\subsection{Deformable Cross Feature Blend (DCFB)}
\subsubsection{Positional Relationship Estimation}
The DCFB Block is a dual-channel input that takes input flow A ($I_A$) and input flow B ($I_B$) as inputs, shown in Fig.~\ref{fig:network}(a). At the Deformable Fusion stage, the two inputs obtain the relative positional deviations (offsets) between $I_A$ and $I_B$ in terms of their corresponding features' absolute positions. To calculate the relative positional deviations between the two images, we introduce depth-wise convolution. Depth-wise convolution partitions the features into groups, prioritizing the calculation of positional deviations within each group and then integrating the positional deviations across multiple groups. This process allows us to obtain the positional deviation of a point in one image with respect to the corresponding point in the other image. The process can be represented as follows:

\begin{figure}[!ht]
	\centering
	\begin{adjustbox}{width= 1.1\textwidth,center}
		\includegraphics{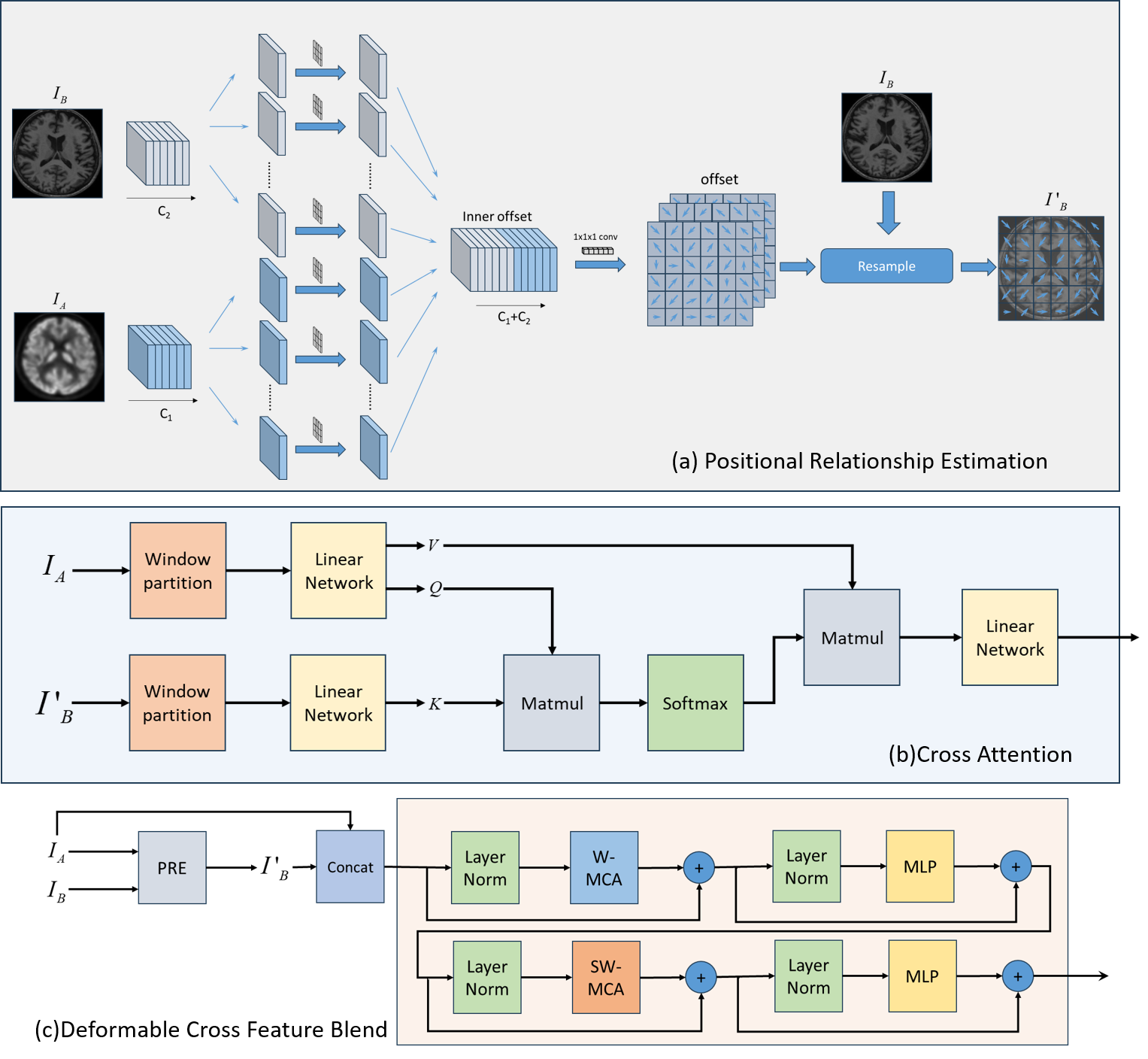} 
	\end{adjustbox}
	\caption{Overall architecture and specific implementation details of the Cross Feature Blend module: (a) PRE is employed to determine the deviation between corresponding points of the two modalities, optimizing the receptive field shape of Cross Attention. (b) The Cross Attention module learns the correlation between the features of the two modalities and fuses their respective characteristics. (c) The Cross Feature Blend module primarily consists of PRE and Deformable Cross Attention.}
	\label{fig:network}
\end{figure}

\begin{equation}
	F_I = Concat(F_{I_A},F_{I_B}) \quad F_{I_A} \in R^{C_1,H,W,D} , F_{I_B} \in R^{C_2,H,W,D} 
\end{equation}
\begin{equation}
	\text{inner offset} = Concat(Conv^1(F^1_{I}),Conv^2(F^2_{I}) ,\dots ,Conv^{C_1+C_2}(F^{C_1+C_2}_{I}))
\end{equation}
\begin{equation}
	offset = Conv_{1\times1\times1}(\text{inner offset})
\end{equation}

Here, $F_{I_A}$ and $F_{I_B}$ are feature maps from $I_A$ and $I_B$, respectively. By concatenating $F_{I_A}$ and $F_{I_B}$, we obtain the feature map $F_I\in \mathbb{R}^{C_1+C_2,H,W,D}$. We employ depth-wise separable convolution with the number of groups set to $C_1+C_2$. Based on the calculation of depth-wise separable convolution, the positional relationships between any point and its $k$-neighborhood (determined by the kernel size of the convolution) can be determined for each feature map. To search for the positional relationships between any two points in a given feature map, we only need to use the $\text{inner offset}$ for computation. Currently, the obtained $\text{inner offset}$ represents only the relative positional deviations within the feature map. To obtain the relative positional deviations between corresponding points across multiple modalities, we apply a $1\times1\times1$ convolution along the channel ($C$) dimension. This allows points $Point_1$ and $Point_2$ in the two feature maps to establish positional relationships based on absolute coordinates, enabling $Point_1$ to locate the position of the corresponding point $Point_1'$ in the other feature map. It is worth noting that $offset$ is a three-channel feature map, where each channel represents the positional deviation in the $x$, $y$, and $z$ directions, respectively.

After obtaining the known positional deviation $offset$, we apply it to $I_B$. $I_B$ is sampled on $offset$ to obtain the position-corrected $I_B'$, expressed as follows:

\begin{equation}
	I_B'=Resample(I_B,offset)
\end{equation}

This means that a point $p$ on $I_B$, after undergoing the deviation $offset_p$, obtains a new coordinate point $p'$, and $p'$ represents the same anatomical location on both $I_A$ and $I_B'$. In other words, the anatomical significance represented by the absolute positions of $I_B'$ and $I_A$ is similar.

\subsubsection{Cross Attention} 

To enable information exchange between $I_A$ and $I_B$, we introduce the Cross Attention module, as shown in Fig.~\ref{fig:network}(b). The Cross Attention module is a component commonly used in computer vision to establish connections between different spatial or channel positions. In our implementation, we adopt the window attention mechanism from Swin Transformer. However, unlike Swin Transformer, which takes a single input, Cross Attention takes the features $F_{I_A}$ and $F_{I_B}$ from two images as input. The two feature maps are initially divided into windows, resulting in $F_{W_A}$ and $F_{W_B}$, respectively. To extract relevant information from $F_{W_A}$ in $F_{W_B}$, we utilize $F_{W_B}$ as the "Key" and $F_{W_A}$ as the "Query." To reconstruct the fused feature map for $I_A$, we employ $F_{W_A}$ as the "Value." The following equations describe the process:

\begin{equation}
Q = F_{W_A} \cdot W_Q \quad K = F_{W_B} \cdot W_K \quad V = F_{W_A} \cdot W_V
\end{equation}

Here, $W_Q$, $W_K$, and $W_V$ represent transformation matrices for the "Query," "Key," and "Value" features, respectively. Finally, the Cross Attention calculation is performed by combining $Q$, $K$, and $V$ as follows:

\begin{equation}
\text{Attention}(Q, K, V) = \text{softmax}\left(\frac{Q \cdot K^T}{\sqrt{d_k}}\right) \cdot V
\end{equation}

Here, $d_k$ denotes the dimension of the key vectors, and $\text{softmax}$ indicates the softmax activation function applied along the dimension of the query. The output of the Cross Attention operation represents the fusion of information from $F_{W_A}$ and $F_{W_B}$, enabling the exchange of relevant information between the two modalities.

\subsubsection{DCFB}
\begin{figure}
	\begin{adjustbox}{width= 0.8\textwidth,center}
		\includegraphics{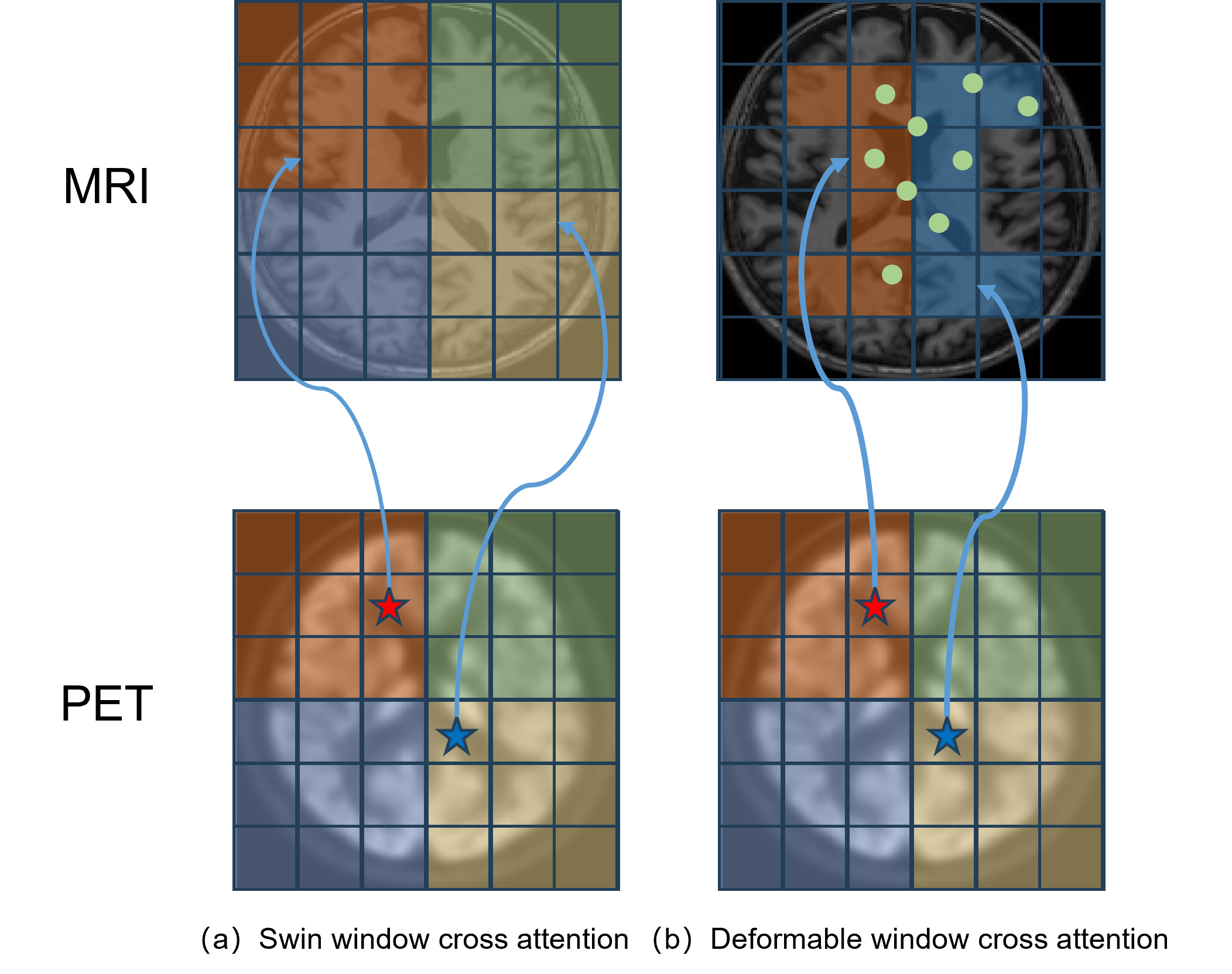} 
	\end{adjustbox}
	\caption{Deformable Window Cross Attention achieves equivalent receptive field deformation through PRE, in contrast to Swin Window Cross Attention.}
 \label{fig:DWCA}
\end{figure}

Although we have applied positional deviation estimation to correct $I_B$, it is important to note that due to the computational characteristics of convolutions and the accumulation of errors, the most effective region for an individual point is its local neighborhood. Beyond this neighborhood, the accuracy of the estimated offset gradually decreases. Therefore, the fundamental role of the offset is to overcome the limited receptive field of the window attention in the Swin Transformer architecture. Moreover, the expansion of the receptive field achieved by the offset follows an irregular pattern. Refer to Fig.~\ref{fig:DWCA} for an illustration.

Assuming $I_A$ represents the features extracted from PET images and $I_B$ represents the features from MRI images, our goal is to query the corresponding points in $I_B$ and their surrounding regions based on $I_A$. However, utilizing the Swin Transformer architecture alone may result in the inability to find corresponding points between the two windows, even with the shift window operation. To address this issue, we propose the Deformable Cross Attention module, which incorporates an offset into the Swin Transformer module. This offset enables the positional correction of $I_B$ beforehand, followed by regular window partitioning on the adjusted $I_B'$. While employing a similar approach as shown in Fig.~\ref{fig:network}(c) for $I_A$ and $I_B'$, the windows on $I_B$ are not regular windows but rather shaped by the features perceived from $I_A$ due to the positional correction applied to $I_B$.

\subsubsection{Loss Functions} 

In our image fusion algorithm, we utilize three distinct loss functions: Structural Similarity Index (SSIM) $\mathcal{L}_{SSIM}$, Normalized Cross-Correlation (NCC) $L_{NCC}$, and L1 loss $L_1$. These loss functions contribute equally to the overall loss with a weight ratio of 1:1:1. Here, we provide a brief explanation of each loss function and their respective contributions to the fusion process. The SSIM ($\mathcal{L}_{SSIM}$) is defined as follows:

\begin{equation}
	\mathcal{L}_{SSIM}(I,J)=\frac{(2\mu_I\mu_J+C_1)(2\sigma_{IJ}+C_2)}{(\mu_I^2+\mu_y^2+C_1)(\sigma_I^2+\sigma_J^2+C_2)}
\end{equation}	
where $I$ and $J$ represent the input and target images, respectively. $\mu$ represents the mean, $\sigma$ represents the standard deviation, and $c1$ and $c2$ are small constants for numerical stability. The Normalized Cross-Correlation ($\mathcal{L}_{NCC}$) is defined as follows:

\begin{equation}
	\mathcal{L}_{NCC}(I, J) = \frac{\sum_{x,y,z} \left(I(x,y,z) - \overline{I}\right) \cdot \left(J(x,y,z) - \overline{J}\right)}{\sqrt{\sum_{x,y,z} \left(I(x,y,z) - \overline{I}\right)^2 \cdot \sum_{x,y,z} \left(J(x,y,z) - \overline{J}\right)^2}}
\end{equation}

where $I$ and $J$ represent the input and target images, respectively. The summation is performed over corresponding image patches. The L1 loss ($\mathcal{L}_{1}$) is defined as follows:

\begin{equation}
	\mathcal{L}_1(I,J)=||I-J||_1
\end{equation}

where $I$ and $J$ represent the input and target images. By combining these three loss functions with equal weights, our image fusion algorithm achieves a balance between multiple objectives. The SSIM loss function focuses on preserving the image structure, the NCC loss function emphasizes structural consistency, and the L1 loss function aims to maintain detail and color consistency. Through the integration of these diverse loss functions, our algorithm comprehensively optimizes the generated image, thereby enhancing the quality of the fusion results.

To avoid biased fusion towards any individual modality solely for the purpose of obtaining favorable loss metrics quickly, we depart from traditional fusion strategies by adopting an end-to-end training approach in our method. To address this concern, we introduce a specific loss function based on the SSIM, which enables us to mitigate the risk of overemphasizing a single modality during the image fusion process solely to optimize loss metrics. This ensures a more balanced fusion outcome and enhances the overall performance of our method:

\begin{equation}
	\mathcal{L}_{pair}=||SSIM(Fusion,MRI)-SSIM(Fusion,PET)||
\end{equation}

The formulation of total loss is as follows:

\begin{equation}
	\begin{aligned}
		\mathcal{L} &= \mathcal{L}_{SSIM}(Fusion,MRI) + \mathcal{L}_{SSIM}(Fusion,PET) \\
		&\phantom{{}=} + \mathcal{L}_{NCC}(Fusion,MRI) + \mathcal{L}_{NCC}(Fusion,PET) \\
		&\phantom{{}=} + \mathcal{L}_{1}(Fusion,MRI) + \mathcal{L}_{1}(Fusion,PET)+ \mathcal{L}_{pair}
	\end{aligned}
\end{equation}

\section{Experiments} 
\subsection{Data Preparation and Evaluation Metrics} 
In this study, we evaluate the performance of our MRI-PET image fusion method on the ADNI-2 dataset\cite{ADNI}. The ADNI-2 dataset consists of 660 participants, each with both MRI and PET images acquired within a maximum time span of three months. We employed the SyN\cite{avants2008symmetric} registration algorithm to align the MRI and PET images to the MNI152 standard space, resulting in images with dimensions $182\times218\times182$. Subsequently, we extracted regions of interest in the form of MRI-PET pairs and resampled them to a size of $128\times128\times128$. Among the collected data, 528 pairs were used as the training set, while 66 pairs were allocated for validation, and another 66 pairs were designated for testing purposes.

To ensure a fair comparison, we conducted a comparative evaluation between our proposed approach and various 2D image fusion methods. For evaluation, we selected the same slice from the MRI, PET, and the fusion image predicted by our method. Objective metrics, SSIM, PSNR, Feature Mutual Information (FMI), and Normalized Mutual Information (NMI), were used to compare the selected slice with alternative methods. This comprehensive assessment enabled us to evaluate the effectiveness and performance of our proposed method against existing approaches in the field of image fusion.

\begin{table}[ht]
	\centering
	\renewcommand{\arraystretch}{1.4} 
	\setlength{\tabcolsep}{5pt} 
	\caption{Quantitative results of the MRI-PET fusion task. The proposed methods demonstrate exceptional performance in terms of the SSIM metric and PSNR when employing 2D-based strategies, establishing a new state-of-the-art benchmark.}
	\begin{tabular}{lccccc}
		\toprule
		Method & 2D-/3D- & $PSNR$ $\uparrow$ & $SSIM$ $\uparrow$ & $NMI$ $\uparrow$ & $FMI$ $\uparrow$ \\
		\midrule
		SwinFuse \cite{Wang2022SwinFuseAR} & 2D & 14.102$\pm$1.490 & 0.623$\pm$0.020 & 1.275$\pm$0.019 & 0.817$\pm$0.009  \\
		MATR \cite{Tang2022MATRMM} & - & 15.997$\pm$1.190 & 0.658$\pm$0.035 & 1.451$\pm$0.009 & 0.795$\pm$0.018\\
		DILRAN \cite{Zhou2022AnAM} & - & 19.028$\pm$0.821 & 0.690$\pm$0.033 & 1.301$\pm$0.015 & 0.806$\pm$0.019 \\
		\midrule
		DC2Fusion(ours)  & 3D & 20.714$\pm$1.377 & 0.718$\pm$0.033 & 1.312$\pm$0.012 & 0.807$\pm$0.020 \\
		\bottomrule
	\end{tabular}
\label{table1}
\end{table}

\subsection{Implementation Details} 
The proposed method was implemented using Pytorch \cite{imambi2021pytorch} on a PC equipped with an NVIDIA TITAN RTX GPU and an NVIDIA RTX A6000 GPU. All models were trained for fewer than 100 epochs using the Adam optimization algorithm, with a learning rate of $1 \times 10^{-4}$ and a batch size of 1. During training, the MR and PET datasets were augmented with random rotation. Our DC2Fusion model consists of three downsampling stages. To handle the limitations imposed by the image size, we employed a window partition strategy with a window size of {2, 2, 2} at each level. Additionally, the number of attention heads employed in each level is {3, 6, 12, 24}.

We conducted a comparative analysis between our method and the 2D image-based fusion methods SwinFuse \cite{Wang2022SwinFuseAR}, MATR \cite{Tang2022MATRMM}, and DILRAN \cite{Zhou2022AnAM}. Each of these methods was evaluated locally using the provided loss functions and hyperparameter settings by their respective authors. MATR \cite{Tang2022MATRMM} and DILRAN \cite{Zhou2022AnAM} methods employ image pair training, where the input images correspond to slices from MRI and PET, respectively. The SwinFuse method \cite{Wang2022SwinFuseAR} does not provide an image pair end-to-end training approach. Therefore, we adapted its fusion process, as described in the paper, which involved fusing infrared and natural images. During testing, a dual-path parallel approach was employed, where MRI and PET slices were separately encoded. Subsequently, a specific fusion strategy (such as the L1 normalization recommended by the authors) was applied to fuse the features of both modalities. Finally, the fused features were passed through the Recon module for reconstruction.

\subsection{Results and Analysis}

\begin{figure}
	\begin{adjustbox}{width= 0.8\textwidth,center}
		\includegraphics{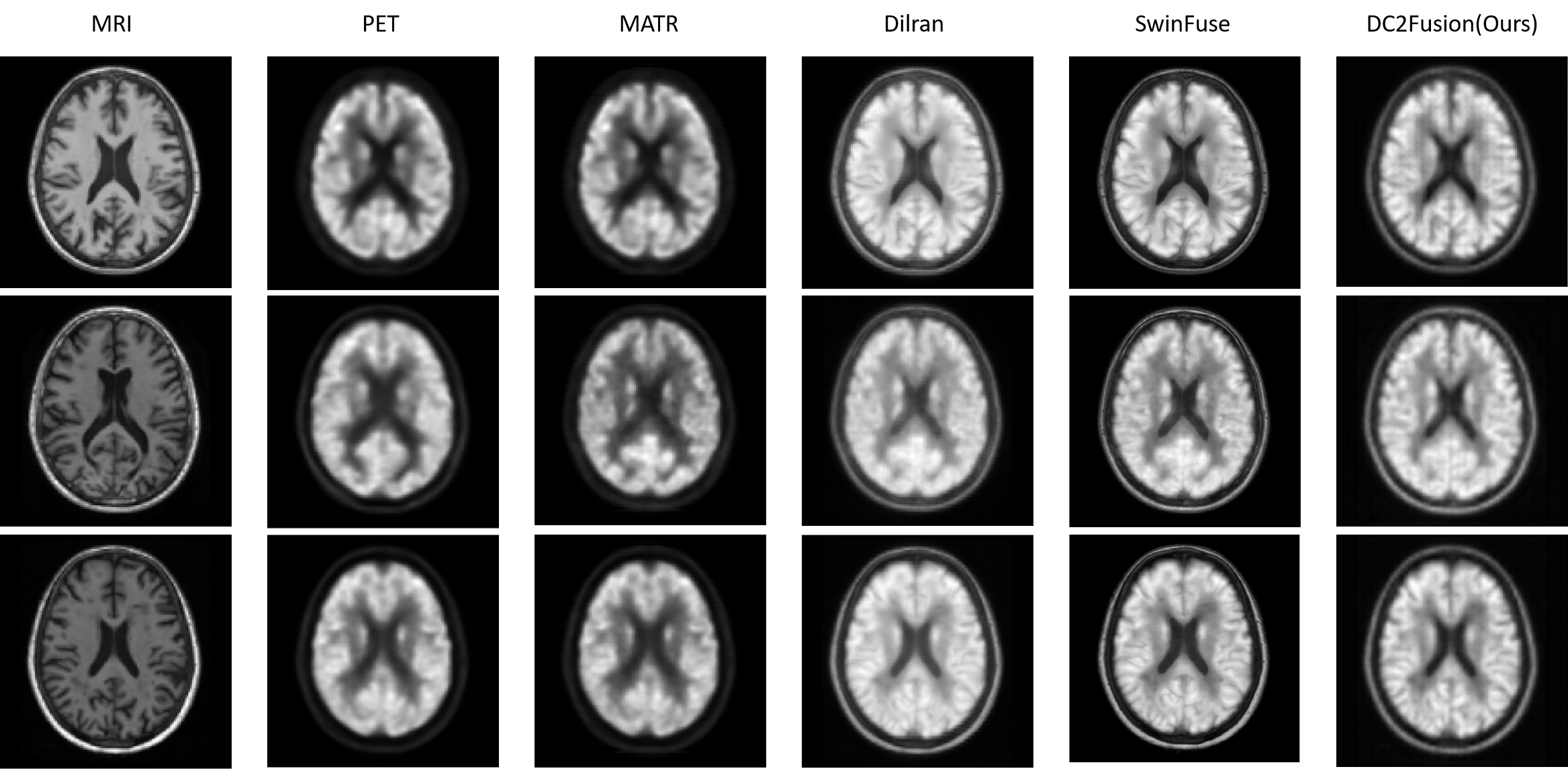} 
	\end{adjustbox}
	\caption{Comparative images of MRI, PET, and other fusion methods on 3 representative PET and MRI image pairs. From left to right: MRI image, PET image, MATR\cite{Tang2022MATRMM}, Dilran\cite{Zhou2022AnAM}, SwinFuse\cite{Wang2022SwinFuseAR}, and DC2Fusion.} \label{Comparetion}
\end{figure}

\begin{figure}
	\begin{adjustbox}{width= 0.8\textwidth,center}
		\includegraphics{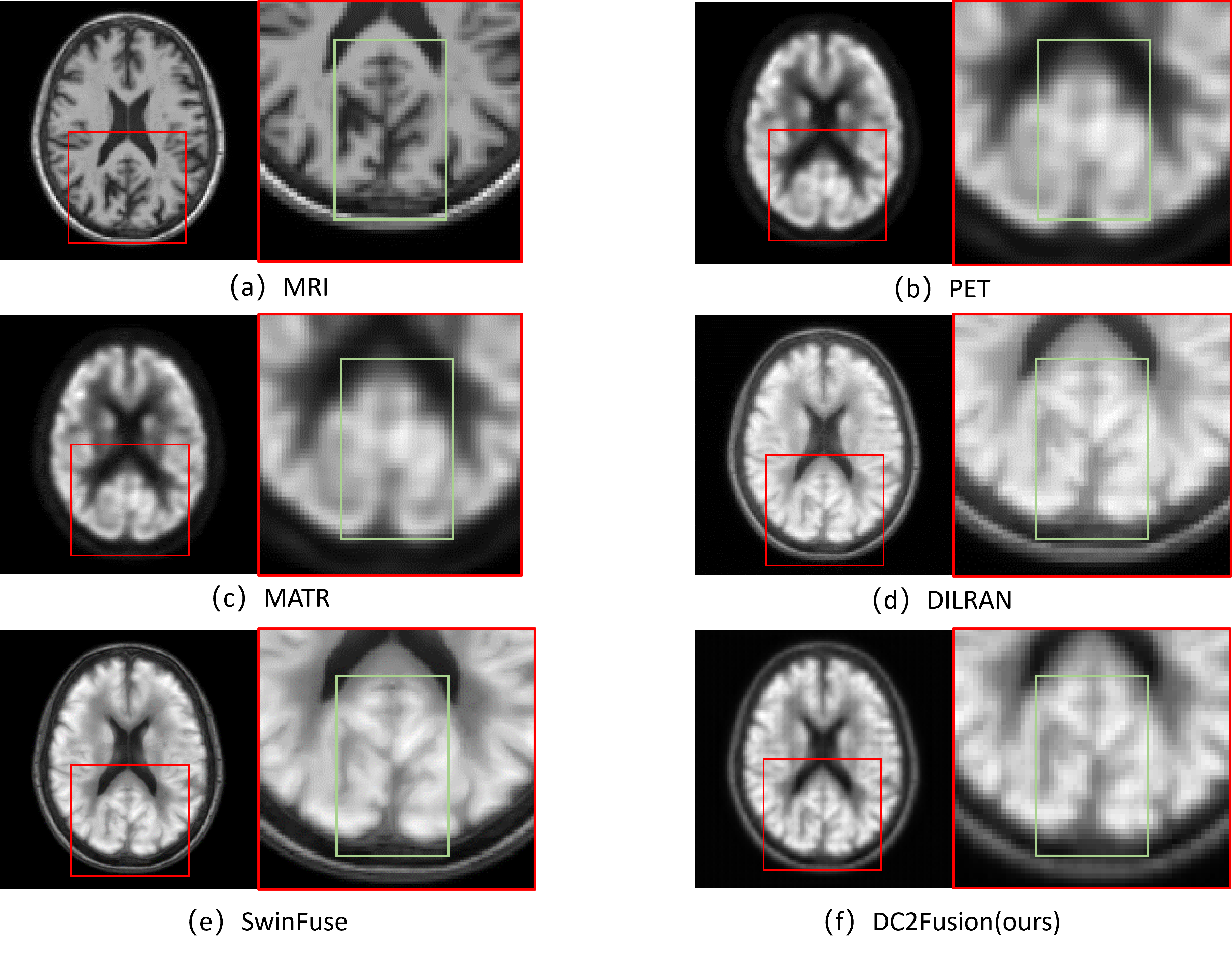} 
	\end{adjustbox}
	\caption{Qualitative comparison of the proposed DC2Fusion with 3 typical and state-of-the-art methods on a representative PET and MRI image pair: (a) MRI image, (b) PET image, (c) MATR\cite{Tang2022MATRMM}, (d) DILRAN\cite{Zhou2022AnAM}, (e) SwinFuse\cite{Wang2022SwinFuseAR}, (f) DC2Fusion.} \label{details}
\end{figure}

Fig.~\ref{Comparetion} illustrates the fusion results of DC2Fusion in comparison with other methods. DILRAN, SwinFuse, and our method all exhibit the ability to preserve the highlighted information from PET while capturing the structural details from MRI. However, the MATR algorithm, which lacks explicit constraints on image fusion during training, demonstrates a complete bias towards learning PET image information and severely lacks the ability to learn from MRI information when applied to the ADNI dataset. Consequently, the model quickly achieves high $SSIM(Fusion, PET)$ values during training, thereby driving the overall loss function gradient. In this experiment, our focus is to compare the fusion image quality of DILRAN, SwinFuse, and DC2Fusion. As shown in Fig.~\ref{Comparetion}, DILRAN exhibits less prominent MRI structural information compared to SwinFuse and DC2Fusion. Moreover, in terms of preserving PET information, SwinFuse and DC2Fusion offer better contrast. For further comparison, Fig.~\ref{details} provides additional details. PET images primarily consist of high-signal information with minimal structural details. DILRAN, SwinFuse, and DC2Fusion effectively fuse the structural information from MRI. In comparison to DC2Fusion, both DILRAN and SwinFuse display slightly thinner structures in the gyri region, which do not entirely match the size of the gyri displayed in the MRI. However, DC2Fusion achieves better alignment with the gyri region of the MRI image. Nevertheless, the structural clarity of DC2Fusion is inferior to SwinFuse. This discrepancy arises because the SwinFuse algorithm does not utilize patch embedding operations or downsampling layers, enabling it to maintain high clarity throughout all layers of the model. However, this approach is limited by GPU memory constraints and cannot be extended to 3D image fusion. Therefore, in our method, we made trade-offs in terms of clarity by introducing techniques such as patch embedding to reduce memory usage and accomplish 3D image fusion tasks.

Table 1 presents a summary of the performance metrics for different methods, including 2D-/3D-fusion, namely PSNR, SSIM, NMI, and FMI. Among the evaluated methods, SwinFuse, a 2D fusion technique, achieved a remarkable FMI score of 0.817. However, it is important to note that FMI solely focuses on feature-based evaluation and does not provide a comprehensive assessment of image fusion quality. SwinFuse obtained lower scores in other metrics, namely PSNR (14.102), SSIM (0.623), and NMI (1.275), suggesting potential limitations in preserving image details, structure, and information content. In contrast, our method, specifically designed for 3D image fusion, exhibited superior performance. DC2Fusion achieved a notable FMI score of 0.807, indicating consistent and correlated features in the fused images. Moreover, DC2Fusion outperformed other methods in terms of PSNR (20.714), SSIM (0.718), and NMI (1.312), highlighting its effectiveness in preserving image details, structural similarity, and information content. Overall, the results demonstrate the competitive performance of our method in medical image fusion, particularly in 3D fusion tasks. These findings emphasize the potential of DC2Fusion in enhancing image quality and preserving structural information, thereby providing valuable insights for further research in the field of medical image fusion.

\begin{figure}
	\begin{adjustbox}{width= 1.0\textwidth,center}
		\includegraphics{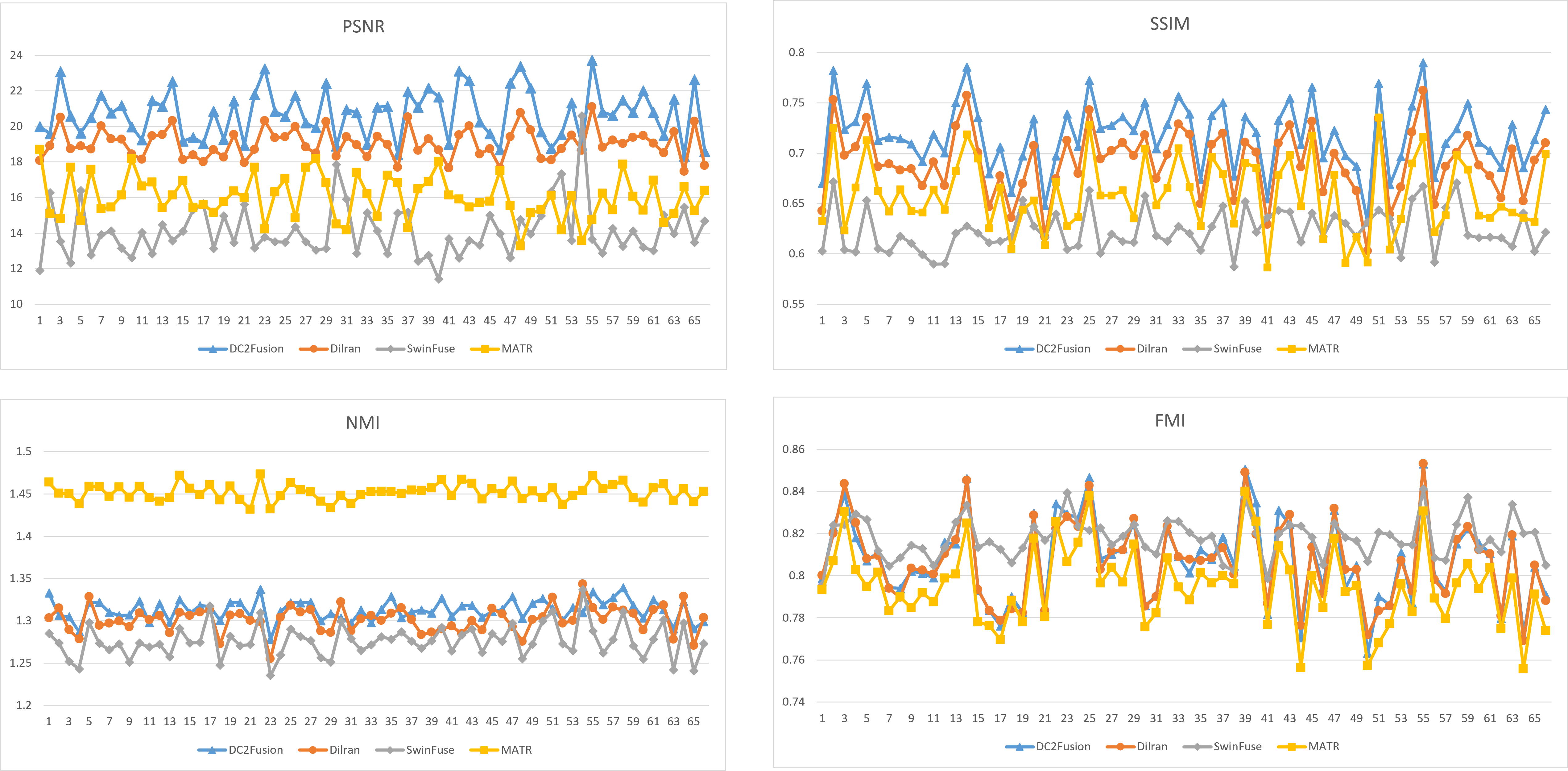} 
	\end{adjustbox}
	\caption{Illustration of the average fusion metrics (PSNR, SSIM, NMI, and FMI) for each sample in our test cases. These metrics are computed by comparing the fusion images with both the MRI and PET images separately. The reported values represent the average scores across all samples.} \label{metrics}
\end{figure}

As shown in Fig.~\ref{metrics}, we present fusion metrics for all samples in our test cases. It is evident that DC2Fusion consistently outperforms other methods in terms of PSNR and SSIM for each sample. However, the NMI metric reveals an anomaly with significantly higher results for MATR. This can be attributed to MATR's fusion results being heavily biased towards the PET modality, resulting in a high consistency with PET images and consequently yielding inflated average values. Nevertheless, these values lack meaningful reference significance. Excluding MATR, our proposed method also achieves better results than other methods in terms of the NMI metric. With regards to the FMI metric, our method does not exhibit a noticeable distinction compared to other methods. Considering Fig.~\ref{details}, although our results may not possess the same level of clarity as other methods, our approach still obtains relatively high FMI scores in terms of these detailed features. This observation demonstrates that our method preserves the feature information during image fusion, even at the cost of reduced clarity.

\section{Conclusion} 
In this study, we propose a novel architecture for 3D MRI-PET image fusion. Our approach, DCFB, achieves cross-modal PRE and provides a deformable solution for enlarging the receptive field during cross-attention. This facilitates effective information exchange between the two modalities. Experimental results demonstrate the outstanding performance of our proposed 3D image fusion method. Although the image fusion quality of our method is currently limited by GPU constraints, resulting in slightly lower clarity compared to 2D fusion, we have provided a novel solution for 3D medical image fusion tasks. Furthermore, we anticipate applying this framework to various modalities in the future, thereby broadening its applicability across different modal combinations.

%
%

\bibliographystyle{splncs04_unsort}
\bibliography{refer}

\end{document}